\documentstyle[epsf]{article}

\newcommand{\beq}{ \begin{equation} }
\newcommand{\eeq}{ \end{equation} }
\newcommand{\bea}{ \begin{eqnarray} }
\newcommand{\eea}{ \end{eqnarray} }

\newcommand{\f}{ \frac }

\newcommand{\ex}{ {\rm e} }

\begin{document}
\thispagestyle{empty}
\parskip=12pt
\raggedbottom

\def\mytoday#1{{ } \ifcase\month \or
 January\or February\or March\or April\or May\or June\or
 July\or August\or September\or October\or November\or December\fi
%\space\number\day ,
 \space \number\year}
\noindent
\hspace*{9cm} BUTP--97/30\\
\vspace*{1cm}
\begin{center}

{\LARGE SU(3) surface tension from the lattice \\ with the fixed point
  action\footnote{Work financially supported by INFN-Italy.}  } 

\vspace{1cm}

Alessandro Papa \\
Institute for Theoretical Physics \\ 
University of Bern \\
Sidlerstrasse 5, CH--3012 Bern, Switzerland

\vspace{0.5cm}

%\mytoday \\ \vspace*{0.5cm}

\begin{center}
 October 1997, revised November 1997
\end{center}
 
\vspace{.5cm}

\nopagebreak[4]

\begin{abstract}
The surface tension at the deconfinement transition of SU(3) is determined
with a parametrized version of the fixed point action of a 
renormalization group transformation on lattices with temporal extent
$N_\tau=3$ and 4 and spatial extent $N_\sigma/N_\tau=3$ and 4. 
A considerable cut-off dependence can be seen in comparison with earlier
determinations from tree level Symanzik and tadpole improved actions.
\end{abstract}

\end{center}

\vspace{0.5cm}


\vfill

\eject

\section{Introduction}
\label{sec:intro}

Lattice Monte Carlo simulations are at present the most effective tool to 
study SU(3) at finite temperature both in the gluon plasma phase, where
perturbation theory is plagued by the well known infrared
problem~\cite{Lin80}, and at the critical temperature $T_c$, i.e. in a  
highly non-perturbative regime. Lattice calculations of the bulk thermodynamic
quantities in the deconfined phase of SU(3) with the Wilson action are
affected, however, by a large cut-off dependence~\cite{BEKLLLP95}, which is
evident also in perturbation theory in the high temperature ideal gas
limit~\cite{BKL96}. The continuum limit can be extrapolated with the Wilson
action only from lattices with temporal extent $N_\tau$ not smaller than
6~\cite{BEKLLLP95}.\footnote{In lattice simulations the temperature $T$ and
  the volume $V$ are determined by the lattice size $N_\sigma^3\times N_\tau $,
  $N_\tau < N_\sigma$, through $T=1/(N_\tau a)$ and $V=(N_\sigma a)^3$.} 
Recent calculations~\cite{BKL96,BKLP97} have revealed that such cut-off
dependence can be drastically suppressed, from temperatures slightly
above $T_c$ onwards, if tree level Symanzik~\cite{Sym83} or tadpole~\cite{LM93}
improved actions are used in numerical simulations. In particular, with the
tadpole improved action the free energy density shows no scaling violation
from $N_\tau=3$ to $N_\tau=4$ and is consistent with the continuum
extrapolated from the Wilson action results. Moreover, the lattice
determinations of the energy density minus three times the pressure from the
tree level Symanzik improved action at $N_\tau=4$ are in agreement with the
continuum. The latter result suggests that the cut-off effects beyond
the tree level order are relatively unimportant in the high temperature phase
of SU(3)~\cite{BKLP97}. In other words, the tree level improvement alone 
can eliminate most of the cut-off dependence. 

Fixed point actions~\cite{HN94} are lattice actions living in the space of
couplings on the straight line which originates at the fixed point (FP) of a
renormalization group (RG) transformation and leaves the critical surface
along the direction of the coupling $g^2$. They are {\em classical} perfect
actions, i.e. their spectral properties are free of cut-off effects 
at the classical level. For this reason and in view of the previous
considerations, FP actions are expected to yield a remarkable improvement in
lattice studies of SU(3) thermodynamics.  

\begin{figure}[htb]
\centering
\leavevmode
\epsfxsize=75mm
\epsfbox{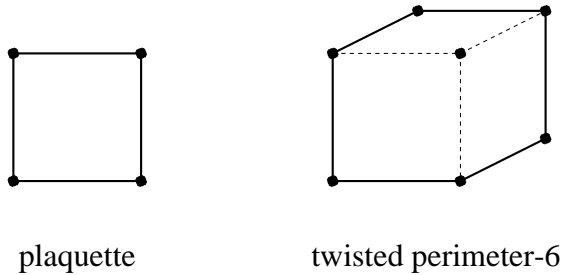}
\caption{Loops considered in the type III parametrized FP action.}
\label{fig:loops}
\end{figure}

Given an arbitrary RG transformation, the FP action of any lattice
configuration is well defined and can be determined numerically by  
multigrid minimization~\cite{HN94,DHHN95a}. For practical reasons, however, in
Monte Carlo simulations only simple enough {\em parametrizations} of the FP
action can be used. Being a potential source of cut-off effects, the
parametrization of a FP action is a very delicate task. In SU(3) lattice gauge
theory, the parametrizations proposed so far are of the form
\beq
S = \beta \; \f{1}{N}\sum_C\;\;\sum_{i\geq1} c_i(C)[N-{\rm
  Re\:Tr}(U_C)]^i \;\;\; , \;\;\;\;\;\;\;\; \beta\equiv\f{2N}{g^2} \;\;\; , 
\eeq 
where the first sum is over all the closed paths $C$ and $U_C$ is the
product of the link variables $U_\mu(n)$ along the path $C$. The couplings
$c_i(C)$ are determined by a fit procedure on the numerically estimated
``exact'' values of the FP action of a representative set of Monte 
Carlo configurations~\cite{DHHN95b,BN96}. 
In Ref.~\cite{BN96} the arbitrariness in the choice of the RG transformation
has been exploited to define a very rotationally symmetric RG transformation,
optimized in order that the related FP action is very short-ranged and,
therefore, easier to be parametrized. This RG transformation has been called
for historical reasons ``type III'' RG transformation. The related FP action
has been parametrized with the plaquette and the twisted perimeter-6 loop (see
Fig.~\ref{fig:loops}), with the couplings $c_i(C)$, $i=1,\ldots,4$, given in
Table~\ref{table:typeIII}.   

In Ref.~\cite{Pap96} the type III parametrized FP action has been used to
determine from the lattice the free energy density of SU(3) at $T/T_c = 4/3, \
3/2, \ 2$. Simulations were performed on lattices as small as $8^3\times 2$ and
$12^3\times 3$. Results at $N_\tau=3$ are in agreement with
the continuum as extrapolated with the Wilson action from lattices with
$N_\tau=6$ and 8~\cite{BEKLLLP95} and with the results from the tadpole
improved action at $N_\tau=3$ and 4~\cite{BKLP97} (for a complete review, see
also~\cite{Lae97}). 

\begin{table}[hbt]
\setlength{\tabcolsep}{1.16pc}
\caption{Couplings of the type III parametrized FP action~\cite{BN96}.}
\label{table:typeIII}
\vspace{.2cm}
\begin{tabular}{lcccc}
\hline
                     & $c_1$     & $c_2$     & $c_3$      & $c_4$   \\
\hline
plaquette            & 0.4822    &  0.2288   &--0.1248    & 0.0228  \\
twisted perimeter-6  & 0.0647    &--0.0224   &  0.0030    & 0.0035  \\
\hline
\end{tabular}
\end{table}

In this paper, the same parametrized FP action has been used to determine 
the surface tension of SU(3) on lattices with temporal extent $N_\tau=3$ and 4
and spatial extent $N_\sigma/N_\tau=3$ and 4. Like the
latent heat, the surface tension is a physical quantity characteristic of the
first order deconfinement transition of SU(3).\footnote{If the high temperature
phase transition of the full QCD were first order, the surface tension would
play an important role in the quark-gluon plasma formation in heavy-ion
experiments and in the nucleation of hadronic matter in the early
universe~\cite{IKKRY94}.} As a consequence, it is affected by the properties
of both the low and the high temperature phase. Since in the latter phase the
use of the type III parametrized FP action allows an impressive reduction of
the cut-off dependence of the thermodynamic quantities, it is interesting to
check if the same occurs for a quantity typical of the critical
region. \newline 
The results for the surface tension obtained in this work have been
then compared with earlier determinations from the Wilson
action~\cite{IKKRY94}, tree level Sy\-man\-zik and tadpole improved 
actions~\cite{BKP96}.  

\section{Monte Carlo results for the surface tension}
\label{sec:res}

At the critical temperature of the first order deconfinement transition of
SU(3), there can be mixed states where the confined and the deconfined phases 
coexist, separated by an interface. These mixed states have an additional free
energy $F = \sigma A$ ($\sigma$ is the surface tension, $A$ the area of the
interface), which indicates that they are less probable than pure states 
of one or the other phase. The frequency distribution of any order parameter
$\Omega$ at the transition has a typical double-peak structure, where the
two peaks correspond to the pure phase configurations, while the region
in-between corresponds to configurations containing an interface. The peaks
become more pronounced when the volume is increased. The probability
distribution of $\Omega$ is given by~\cite{IKKRY94}
\bea
P(\Omega) & = & c_1 \ex^{-f_1 V/T} \ex^{-(\Omega-\Omega_1)^2/d_1^2}
            +   c_2 \ex^{-f_2 V/T} \ex^{-(\Omega-\Omega_2)^2/d_2^2}\nonumber \\
          & + & c_m \ex^{-(f_1 V_1 + f_2V_2 + 2\sigma A)/T}\;\;\;, \\
& & c_i \propto V^{1/2}, \;\;\; d_i \propto V^{-1/2}, \;\;\;\;\; i=1,2\;\;\;,
 \nonumber 
\label{eq:Pdistr}
\eea
where $f_1$, $f_2$ are the free energy densities of the two pure phases and
$V_1$, $V_2$ are their volumes. The factor 2 in the free energy of the
interface  appears since at finite volume with periodic boundary conditions
two interfaces are needed to separate two volumes. In the previous formula,
the dependence on $\Omega$ enters through the relation $ \Omega = (V_1\Omega_1
+ V_2\Omega_2)/V$.
At $T_c$ and at infinite volume, the free energy densities in both phases are
identical, $f_1=f_2$. The leading volume dependence of the 
surface tension is determined by~\cite{IKKRY94}
\beq
\left(\f{\sigma}{T_c^3}\right)_V = - \f{1}{2} \left(\f{N_\tau}{N_{\sigma}}
\right)^2 
\ln\left(\f{P_{\rm min}}{P_{\rm max,1}^{\gamma_1}\:P_{\rm
      max,2}^{\gamma_2}}\right) \;\;\; ,  
\label{eq:s}
\eeq
where $P_{\rm min}$ is the minimum of the distribution
$P(\Omega)$, $P_{\rm max,1}$ and $P_{\rm max,2}$ are the two maxima
corresponding to the values $\Omega_1$ and $\Omega_2$ of $\Omega$ in the pure
states of the two phases at infinite volume. The weights $\gamma_1$
and $\gamma_2$ are determined imposing $ \langle \Omega \rangle = \gamma_1
\Omega_1 + \gamma_2 \Omega_2$, with $\gamma_1 + \gamma_2 = 1$.

\begin{table}[tb]
\setlength{\tabcolsep}{0.44pc}
\caption{Parameters of the runs and results for $\beta_c$ and
$\chi_L/N_\sigma^3$. Errors have been estimated by the
jackknife method.}
\label{table:s}
\vspace{.2cm}
\centering
\begin{tabular}{lccccc}
\hline
lattice & \# $\beta$ & \# iterations & $\beta$ reweighting & $\beta_c$ &
$\chi_L/N_\sigma^3$   \\
\hline
$12^3\times3$ &3& 645099 & 3.58995 & 3.58982(9) & $2.958(20)\times10^{-2}$ \\
$12^3\times4$ &1& 171209 & 3.69915 & 3.7007(3)  & $1.013(16)\times10^{-2}$ \\ 
$16^3\times4$ &3& 135662 & 3.70025 & 3.7009(5)  & $8.87(20)\times10^{-3}$  \\ 
\hline
\end{tabular}
\end{table}

\begin{figure}[htb]

\centering
\leavevmode
\epsfxsize=100mm
\epsfbox{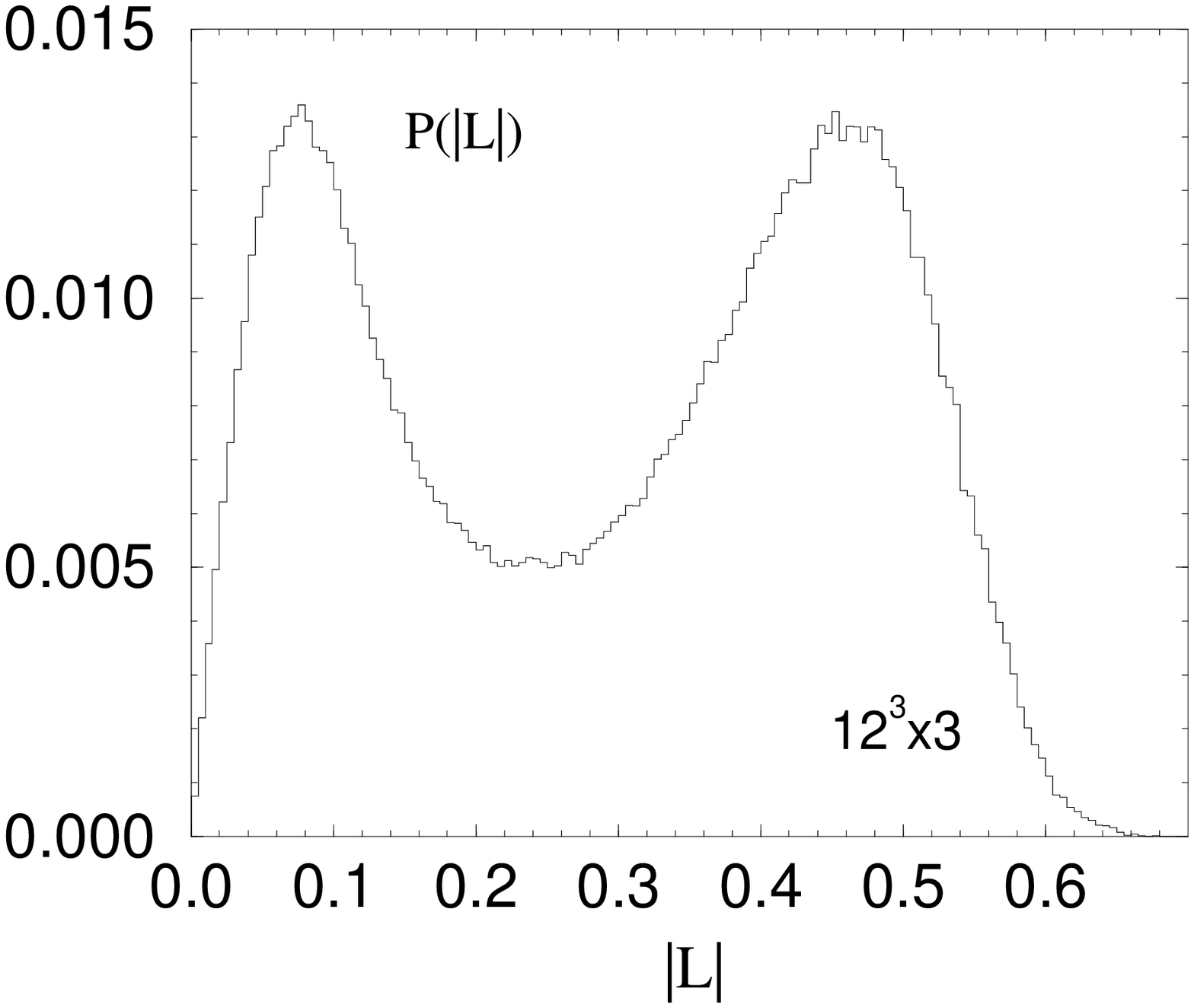}

\centering
\leavevmode
\epsfxsize=100mm
\epsfbox{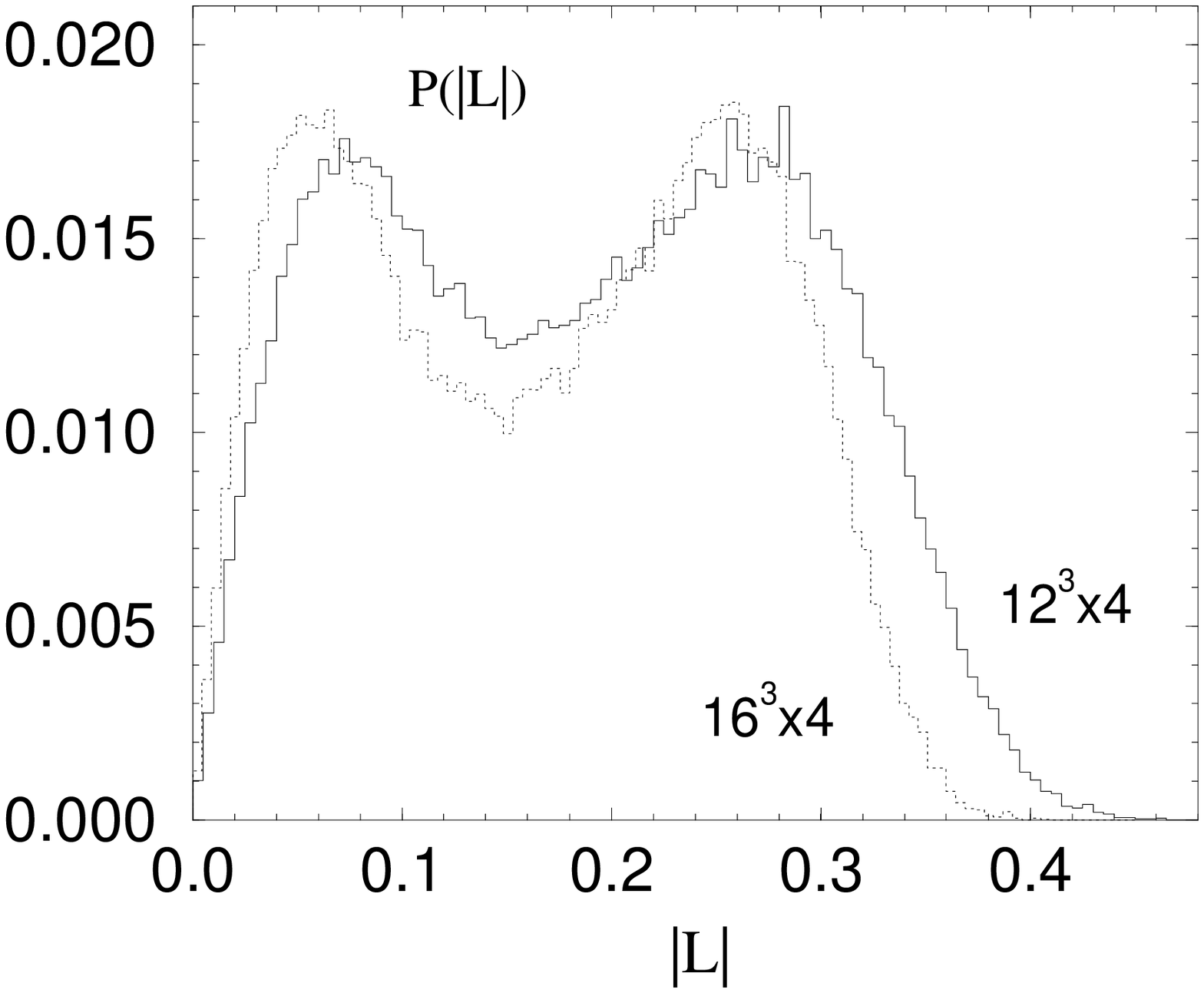}

\caption{Polyakov loop distribution on the $12^3\times3$ lattice and on the
  $12^3\times4$ and $16^3\times4$ lattices. Data on the different lattices
  have been normalized to the corresponding total statistics.}
\label{fig:P(|L|)}

\end{figure}

A convenient choice for the order parameter in SU(3) is the absolute value of
the Polyakov loop $L = 1/N_\sigma^3 \sum_{\vec{n}}{\rm Tr}
\prod_{n_4=1}^{N_\tau} U_\mu (\vec{n},n_4)$. Numerical simulations  
have been performed on three lattices at values of $\beta$ close to
criticality. A summary of the parameters of the runs is given
in Table~\ref{table:s}. Here one iteration means the combination of a 20-hit
Metropolis and 4 over-relaxation updatings; the runs have been done on
DEC-alpha machines. In all the numerical simulations frequent flips
of the order parameter $|L|$ could be observed between the two phases, thus
indicating a proper sampling of the thermal ensemble at criticality. The
occurrence of frequent jumps was expected in view of the relatively low
$N_\sigma/N_\tau$ ratio. The critical couplings have been determined through
the location of the peak in the Polyakov loop susceptibility $\chi_L =
N_\sigma^3 (\langle |L|^2 \rangle - \langle |L|\rangle^2)$. They are in
agreement within errors with the results of Ref.~\cite{BN96} where the so
called Columbia definition was adopted.\footnote{This comparison is important
  since in Ref.~\cite{Pap96} the physical scale in the determination of the
  free energy density was set using the critical couplings as obtained in
  Ref.~\cite{BN96}.}  
The Polyakov loop distribution on the $12^3\times3$ lattice and on the two
lattices with $N_\tau=4$ is presented in Fig.~\ref{fig:P(|L|)}. Data at
different $\beta$ on the same lattice have been interpolated by
Ferrenberg-Swendsen reweighting in order to make the peaks in the Polyakov
loop distribution have the same height. In order to determine $P_{\rm min}$,
$P_{\rm max,1}$ and $P_{\rm max,2}$, the statistical fluctuations in the
vicinity of the extrema of $P(|L|)$ have been smoothed out by a polynomial fit.

In Table~\ref{table:s_sum} the results of this work for $\sigma/T_c^3$ are
compared with other determinations from the Wilson action~\cite{IKKRY94}, the
tree level Symanzik improved action (plaquette + $(1\times2)$ loop) and the 
tadpole improved action (with the same loops)~\cite{BKP96}. The infinite
volume extrapolations have been done according to the ansatz~\cite{IKKRY94}  
\beq
\left(\f{\sigma}{T_c^3}\right)_V = \left(\f{\sigma}{T_c^3}\right)
- \left(\f{N_\tau}{N_{\sigma}}\right)^2 \left[c +
\f{1}{4}\ln N_\sigma\right].
\eeq
The result on the $12^3\times3$ lattice and the infinite volume extrapolation
at $N_\tau=4$ for the tadpole improved action indicate no cut-off dependence
and a continuum value of $\sigma/T_c^3$ equal to 0.0155(16) (corresponding to
$\sigma\sim 7 \ {\rm MeV/fm}^2$).\footnote{Here and in the following it is
  assumed that the infinite volume extrapolation at $N_\tau=3$ does not differ
  too much from the result on the $12^3\times3$ lattice.} Moreover, this
continuum value is in agreement with the infinite volume extrapolation at
$N_\tau=4$ from the tree level Symanzik improved action and with the
extrapolation in $1/N_\tau^2$ to $N_\tau=\infty$ of the results with the
Wilson action at $N_\tau=4$ and 6. \newline  
In the case of the type III parametrized FP action, the result on the 
$12^3\times3$ lattice and the infinite volume extrapolation at $N_\tau=4$ are
consistent with scaling, but at a value considerably larger than the continuum
determined from the tadpole improved action. If one observes, however, 
that the infinite volume extrapolation at $N_\tau=4$ is probably not reliable
for the smallness of the lattices involved and assumes that the true
extrapolated value cannot be too different from the result on the
$16^3\times4$ lattice, then it must be concluded that there is a large scaling
violation. Although the result on the $16^3\times4$ lattice is not inconsistent
with the corresponding determinations from the tree level Symanzik and the
tadpole actions, the result on the $12^3\times3$ lattice differs from the
continuum more than in the case of the tree level Symanzik action. 

\begin{table}[bt]
\setlength{\tabcolsep}{0.30pc}
\centering
\caption{$\sigma/T_c^3$ for various lattice actions on several
  lattices. Column 7 gives the results for $\gamma_1$ from the type
  III parametrized FP action. Errors have been estimated by the
  jackknife method.} 
\label{table:s_sum}
\vspace{.2cm}
\begin{tabular}{ccccccc}
\hline
&
& \multicolumn{1}{c}{Wilson~\cite{IKKRY94}}
& \multicolumn{1}{c}{tree level~\cite{BKP96}} 
& \multicolumn{1}{c}{tadpole~\cite{BKP96}} 
& \multicolumn{2}{c}{type III FP} \\
\cline{6 - 7}
\multicolumn{1}{c}{$N_\tau$} 
& \multicolumn{1}{c}{volume} 
& \multicolumn{1}{c}{$\sigma/T_c^3$} 
& \multicolumn{1}{c}{$\sigma/T_c^3$} 
& \multicolumn{1}{c}{$\sigma/T_c^3$} 
& \multicolumn{1}{c}{$\sigma/T_c^3$} 
& \multicolumn{1}{c}{$\gamma_1$} \\
\hline
 3 & $12^3$         &            & 0.0234(24) & 0.0158(11) & 0.0307(8)  &
 0.420(10) \\
\hline
 4 & $12^2\times24$ & 0.0241(27) &            &            &            & \\
 4 & $24^2\times36$ & 0.0300(16) &            &            &            & \\ 
 4 & $12^3$         &            &            &            & 0.0196(11) &
 0.409(21) \\
 4 & $16^3$         &            & 0.0148(16) & 0.0147(14) & 0.0180(21) & 
 0.441(23) \\
 4 & $24^3$         &            & 0.0136(25) & 0.0119(21) &            & \\
 4 & $32^3$         &            & 0.0116(23) & 0.0125(17) &            & \\
\hline
 4 & $\infty$       & 0.0295(21) & 0.0152(26) & 0.0152(20) & 0.026(5)   & \\
\hline
 6 & $20^3$         & 0.0123(28) &            &            &            & \\
 6 & $24^3$         & 0.0143(22) &            &            &            & \\
 6 & $36^2\times48$ & 0.0164(26) &            &            &            & \\
\hline
 6 & $\infty$       & 0.0218(33) &            &            &            & \\
\hline
\end{tabular}
\end{table}

This scenario would suggest that the type III parametrized FP action does not
improve at $T=T_c$, at least as far as the surface tension is concerned,
although it clearly does, from temperatures slightly above $T_c$ onwards, in
the case of the free energy density~\cite{Pap96}.  

Before accepting this conclusion, possible sources of error have been
investigated. The procedure followed to find the results of this work has
been verified by re-calculating some old results for the surface tension at
$N_\tau=2$ obtained with the Wilson action~\cite{JBK92}. 

The use of the {\em na\"{\i}ve} Polyakov loop as order parameter, instead
of the corresponding classically perfect operator, could represent in
principle a source of lattice artifacts. This possibility has to be excluded,
however, for several reasons: first of all, the type III parametrized FP
action was optimized in Ref.~\cite{BN96} in order to dump the violation of
rotational symmetry in the correlators of na\"{\i}ve Polyakov loops and,
consequently, in the static $q\bar q$ potential. The tests performed in
Ref.~\cite{BN96} provided impressively good results, in comparison with 
previous parametrizations of the FP action. Moreover, the surface tension
depends only on the height of the minimum and of the maxima of the
distribution $P(|L|)$, while possible lattice artifacts in the corresponding
values of $|L|$ should play a minor role. This statement is supported by the
fact that the nice results for the surface tension from the tadpole improved
action of Ref.~\cite{BKP96} were obtained with the na\"{\i}ve Polyakov
loop. Finally, it has been explicitly checked in the present work that using
the FP action density itself as order parameter instead of the absolute value
of the Polyakov loop does not change the scenario, although the statistical
fluctuations are larger. 

The most natural explanation of the large observed cut-off dependence is that 
it is induced by the parametrization of the FP action, which perhaps
fails to represent the exact FP action on configurations typical of the
critical region. By comparing the results from the type III parametrized FP
action with those of Ref.~\cite{BKP96}, one could also argue that planar
(although less local) loops are preferable in lattice actions for
simulations in presence of surfaces. The fact that the type III parametrized
FP action does not satisfy the Symanzik condition at the tree
level~\cite{LW85} could play a role in this respect. 

It must be stressed, however, that the surface tension is a difficult quantity
to determine for the high statistics required and for the complicated volume
dependence. Further checks are necessary to confirm the previous conclusion. 

\section{Acknowledgments} 

I thank F.~Niedermayer and P.~Hasenfratz for several stimulating
discussions. This work was financially supported by a grant of INFN-Italy.

\end{document}